\documentclass{optica-article}

\journal{opticajournal} 

\articletype{Research Article}

\usepackage{lineno}

\begin{document}

\title{Efficient Optimisation of Physical Reservoir Computers using only a Delayed Input}

\author{Enrico Picco\authormark{1,*}, Lina Jaurigue\authormark{2}, Kathy L{\"u}dge\authormark{2} and Serge Massar\authormark{1}}

\address{\authormark{1} Laboratoire d’Information Quantique CP224, Université libre de Bruxelles (ULB), Av. F. D. Roosevelt 50, 1050 Bruxelles, Belgium\\
\authormark{2} Institute of Physics, Technische Universität Ilmenau, Weimarer Straße 25, 98693 Ilmenau, Germany
}

\email{\authormark{*}enrico.picco@ulb.be} 


\begin{abstract*} 
We present an experimental validation of a recently proposed optimization technique for reservoir computing, using an optoelectronic setup. Reservoir computing is a robust framework for signal processing applications, and the development of efficient optimization approaches remains a key challenge.  The technique we address leverages solely a delayed version of the input signal to identify the optimal operational region of the reservoir, simplifying the traditionally time-consuming task of hyperparameter tuning. We verify the effectiveness of this approach on different benchmark tasks and reservoir operating conditions.


\end{abstract*}

\section{Introduction}


In recent years, the quest for efficient and powerful machine learning paradigms has led researchers to explore unconventional computational architectures inspired by the dynamics of physical systems. 
Reservoir Computing (RC) \cite{jaeger2004harnessing} is a possible promising approach. 
RC diverges from conventional neural networks by introducing a distinct architecture characterized by a fixed, randomly initialized recurrent layer, called "reservoir", coupled with a simple linear readout layer: its characteristic structure reduces the training computational complexity, and consequently, the energy consumption. 
RC possesses a remarkable capacity to process temporal data, distinguishing itself across a large variety of tasks, such as equalization of distorted nonlinear communication channels \cite{jaeger2004harnessing, estebanez202256}, audio processing \cite{verstraeten2005isolated, triefenbach2010phoneme}, weather prediction \cite{mammedov2022weather}, image \cite{schaetti2016echo, antonik2019large} and video \cite{antonik2019human, picco2023high} classification. RC has also gathered attention for its feasibility to being implemented on a wide range of physical substrates \cite{tanaka2019recent, nakajima2021reservoir}. In particular, photonic RC stands out as one of the most promising hardware platforms, thanks to its advantages in parallelization \cite{liutkus2014imaging, rafayelyan2020large}, high speed \cite{larger2017high} and minimal hardware requirements \cite{van2017advances}.

Hyperparameter optimisation is a crucial aspect of tuning reservoir computers, aiming to enhance their performance across various tasks. Various optimization techniques are employed to search for the most effective hyperparameter values. Grid search, random search, and more sophisticated methods like Bayesian optimization \cite{yperman2016bayesian, antonik2021bayesian} and genetic algorithms \cite{FERREIRA20134172, basterrech2022re} have been proposed. Nonetheless, the optimization of hyperparameters remains a critical challenge to the widespread adoption of RC \cite{joy2022rctorch}.
The challenge lies in exploring the vast hyperparameter space to find configurations that yield optimal results for a given task, and the sensitivity of the reservoir to hyperparameters. The high-dimensional nature of the hyperparameter space can make the search process computationally expensive and time-consuming, even more so in the case of slow experimental reservoirs that require more hyperparameters and for which one iteration of the grid search takes up to several hours \cite{antonik2019human}. 

In this work we study a novel approach, first introduced in \cite{Jaurige2021reservoir} in numerical simulations, based on the use of a delayed input. In \cite{Jaurige2021reservoir} and later in \cite{Jaurige2024reducing}, Jaurige at al. show that adding a time-delayed version of the input can considerably improve the performance of a reservoir whose hyperparameters are not adjusted. This powerful approach requires to tune only two parameters instead of the many hyperparameters usually associated to a reservoir, thus allowing the optimisation with little computational cost. More recently, the use of time delayed inputs was investigated numerically in \cite{castro2023wavelength} in the context of a RC based on silicon microrings.

We test this approach for the first time on an experimental RC.
We use the architecture based on time multiplexing first introduced in \cite{appeltant2011information}, but in an optoelectronic version as introduced in \cite{paquot2012optoelectronic} and since studied extensively,  see e.g. \cite{antonik2016online, hermans2016embodiment, nakajima2022physical}.
We assess the impact of this technique on different tasks involving time-series forecasting and audio classification, to verify its effectiveness using data of various nature with different intrinsic timescales. We show the superiority, in terms of performance, of this procedure versus the standard hyperparameters optimisation. 

We hope with this work to introduce a new technique in RC community for the optimisation of physical reservoir computers; 
as discussed in the conclusion, this approach could be particularly relevant for hardware reservoir computing where it may not be possible to optimise all hyperparameters, or the hyperparameters may drift over time.

\section{Time-delay Reservoir Computing}
\label{sec_RC}

A reservoir computer consists of three main components: an input layer, a reservoir and an output layer. The input layer maps the input signal into the reservoir. The reservoir is a recurrent neural network with fixed interconnections. 
The reservoir layer exploits the rich dynamics of its internal state, acting as a complex computational substrate that temporally transforms input signals into high-dimensional representations. During the training phase, the reservoir's states are used to evaluate the parameters of the linear output layer.


In this work, we use the well-known delay-based reservoir computer based
on a single nonlinear node. The reader can refer to \cite{hulser2022role, van2017advances} for reviews on time-delay RC.
In this implementation, the reservoir nodes are arranged in a topology similar to a ring structure where each node is connected to its neighboring nodes.
The dynamics of a time-delay reservoir of size $N$ in discrete time is described by:
\begin{equation}
\label{eq_T-M}
\begin{aligned}
& x_0(n+1) = f(\alpha x_{N-1}(n-1) + \beta M_0u(n)) \\
& x_i(n+1) = f(\alpha x_{i-1}(n) + \beta M_iu(n)) & i = 1,.., N-1,
\end{aligned}
\end{equation}
and the reservoir's output $y$ is evaluated through a simple linear operation $y(n) = W_{out} x(n)$ and it is used to evaluate the reservoir computer's performance, selecting the task-specific figure or merit. 
$x(n)$ is the $N$-size vector of the reservoir states at timestep  $n$; $u(n)$ is the temporal dependent input signal of size $K$; $f$ is a nonlinear activation function, in our experiments is a sinusoidal function (cf. sec. \ref{sec_setup});  $\alpha$ is the feedback attenuation, which represents the strength of the interconnections between reservoir's nodes; the input mask $M$ maps the input signal in the reservoir, and its coefficient are usually drawn at random from a uniform distribution in the range $[-1, +1]$; $\beta$ is the scaling of the input signal, usually called input strength.  
After running the reservoir, all the state vectors $x(n)$ are collected in a $N$x$K$ state matrix $X$ which is used for the training, i.e. the to obtain the $N$ weights $W_{out}$ of the linear readout layer. In this work we use the regularized linear regression [43]:
\begin{equation}
W_{out} = (X^TX+\lambda I)^{-1}X^T \tilde y,
\label{eq_ridge}
\end{equation}
where $\tilde y$ is the target signal and $\lambda$ is the regularization parameter. In the case of a classification task with $C$ output classes, $W_{out}$ becomes a $C$x$N$ matrix.


\section{Reservoir Computing with Delayed Input}
The idea of this work is to used a delayed version of the input time-series $u(n)$ as the input signal to the reservoir. We use two $N$x$K$ input masks $M_1$ and $M_2$ with values sampled from uniform distributions in the range $[0,1]$. The value of the new input $J$ at timestep $n$ is then defined as:
\begin{equation}
\label{eq_dly_inp}
J(n) = \beta_{1}u(n)M_1 (n) + \beta_{2}u(n-d)M_2 (n) + J_0,
\end{equation}
where $d$ is the delay, $\beta_{1}$ and $\beta_{2}$ are scaling parameters, and $J_0$ is a constant bias. 
The dynamics of the reservoir of equation (\ref{eq_T-M}) become then:
\begin{equation} 
\label{eq_T-M_dly}
\begin{aligned}
& x_0(n+1) = f(\alpha x_{N-1}(n-1) + J(n+1)) \\
& x_i(n+1) = f(\alpha x_{i-1}(n) + J(n+1)) & i = 1,.., N-1.
\end{aligned}
\end{equation}
The new input $J$ contains two different components: one related to the input at the present timestep, and one related to the input value $d$ timesteps in the past. The masking process is now made with the two different masks $M_1$ and $M_2$.
The idea is to optimise the reservoir tuning only the values of $\beta_2$ and $d$, while keeping fixed the parameters specific to the reservoir (such as $\alpha$ and $\beta_1$). A detailed explanation of this procedure can be found in section \ref{sec_results}.


\section{Experimental Setup}
\label{sec_setup}

\begin{figure}[h!]
\centering\includegraphics[width=7cm]{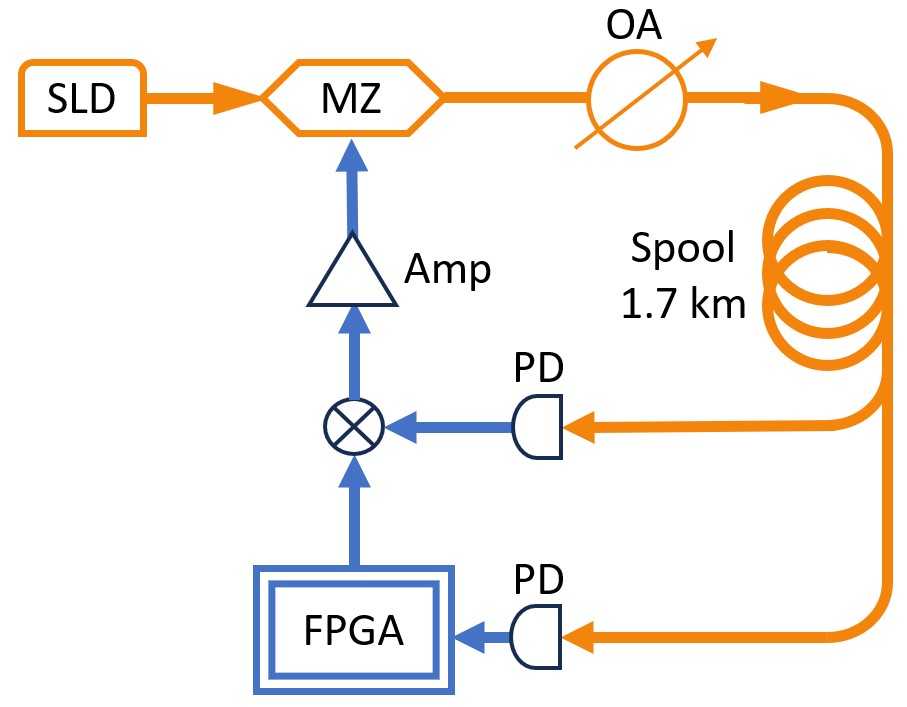}
\caption{Optoelectronic reservoir computer. In orange the optic part and in blue the electronic part. SLD: superluminescent diode. MZ: Mach-Zender intensity modulator. OA: Optical Attenuator. PD: Photo-Detector. FPGA: Field Programmable Gate Array. Amp: amplifier. }
\label{fig_setup}
\end{figure}

Our optoelectronic setup, shown in Figure \ref{fig_setup}, implements the time-delay RC introduced in section \ref{sec_RC} and it is similar to experimental systems introduced in previous works \cite{antonik2016online, hermans2016embodiment}.
A superluminescent diode (Thorlabs SLD1550P-A40) generates broadband light at 1550 nm.
A Field Programmable Gate Array (FPGA) board generates the reservoir's input signal. This electrical signal coming from the FPGA drives an electro-optic Mach-Zender (MZ) intensity modulator (EOSPACE AX2X2-0MSS-12): 
the input signal is thus represented in the optical domain. Moreover, the sinusoidal transfer function of the MZ modulator works in intensity only, so that the MZ acts as
$I_0 sin^2 (V + \frac{\pi}{4}) = I_0 ( \frac{1}{2} + \frac{1}{2} sin(2V) )$. Because the photodetector and the amplifier have a low pass filter, the constant term has no role in the dynamics, leading to the equations (\ref{eq_T-M}), with $f$ as a sinusoidal nonlinearity. The MZ modulator takes roughly 5 ns to reach a steady state response, while the mask step duration is in the order of hundreds of ns: it is therefore possible to describe the system using time discrete equations.

The light passes trough an Optical Attenuator (JDS HA9) which attenuates the light intensity in the loop by a fixed configurable factor. Then, the light travels in a fiber spool, whose length of 1.7 km corresponds to a delay of 7.94 $\mu$s. This is the delay used to implement the delay-based RC: at this point, after the spool, the light represents the time-multiplexed values of reservoir's nodes. Part of this light is collected by a Photo-Detector (PD), electrically summed to the input coming from the FPGA, and amplified (with a Mini Circuits ZHL-32A+ coaxial amplifier) to drive the MZ modulator over its full $V_{\pi}$ range . The other part of the light is collected by another PD, stored by the FPGA and saved on a PC for offline computations of the output weights. 



\section{Tasks}
\label{sec_tasks}

We test our optoelectronic reservoir computer on four different tasks: NARMA10, Mackey-Glass system, Spoken Digit Recognition and Speaker Recognition. These are all widely used benchmark tasks in RC community. The first two are time-series predictions tasks, whose goal is to forecast future values in a sequence based on historical information; the last two are classification tasks, aiming to accurately identify and categorize audio-based temporal signals. 

\subsection{NARMA10}
The NARMA10 task is a widely used benchmark in RC, and involves predicting the next value in a sequence generated by a NARMA  (Nonlinear Auto-Regressive Moving Average) process of order 10. This task evaluates the capability of a RC to capture complex temporal dependencies and make accurate predictions in nonlinear dynamic systems.
The input signal $u(n)$ is randomly drawn from a uniform distribution between $[0, 0.5]$. The output $q(n)$ of the NARMA10 system is defined as:
\begin{equation}
\label{eq_narma}
q(n+1) = 0.3q(n)+0.05q(n)(\sum_{i=0}^{9}q(n-1))+1.5u(n-9)u(n)+0.1.
\end{equation}
The aim is, given $u(n)$, to predict $q(n)$.
We evaluate the accuracy of the system using the normalised mean square error (NMSE), defined as 
\begin{equation}
\label{eq_NMSE}
NMSE = \frac{\langle (y(n)-\tilde{y}(n))^2 \rangle}{\langle \tilde{y}(n)- \langle \tilde{y}(n) \rangle)^2 \rangle}, 
\end{equation}
where $y(n)$ is the reservoir's output and  $\tilde{y}(n)$ is the target signal ($\tilde{y}(n)=q(n)$ for NARMA10).

\subsection{Mackey-Glass system}
The Mackey-Glass system describes a time-delay differential equation that exhibits chaotic behavior \cite{Mackey1997}:
\begin{equation}
\label{eq_MG}
\frac{dx}{dt}= \beta \frac{x(t-\tau)}{1+x(t-\tau)^{n}} - \gamma x.
\end{equation}
With $\beta=0.2$, $\tau=17$, $n=10$, $\gamma=0.1$ and timestep $dt=1$. In this work we try to forecast the values of the series ten time steps into the future. Similarly to NARMA10, also for this task we use the NMSE (eq. \ref{eq_NMSE}) as a figure of merit.

\subsection{Spoken Digit Recognition}
The Spoken Digit Recognition dataset consists of 500 total utterances of the 10 (from 0 to 9) digits. Each digit is repeated ten times by five different individuals. We use a dataset with a 3 dB Signal-To-Noise (SNR) ratio babble noise, to increase the difficulty of the task and to evaluate the system's performance in noisy environments. The dataset is split 450/50 for train/test.
The audio signals are pre-processed using the Lyon Passive Ear model \cite{lyon1982computational}. This model emulates the auditory canal's biological response and transforms the original time-domain utterances in a 86-channels frequency representation, which is used as input signal. The length of the utterances ranges from 22 to 95 timesteps, with an average length of approximately 68 timesteps each.

In order to address the small size of our dataset containing 500 digits, we implement k-fold cross-validation with k=10. This technique involves dividing the dataset into 10 equal parts of 50 digits each. The training process is then repeated 10 times, where each time a different subset of 50 digits is used for testing while the remaining 450 are used for training.
Since this is is a classification task with 10 output classes, we use 10 distinct linear classifiers: each classifier is designed to output "+1" if the input matches the corresponding digit and "-1" otherwise.
The system selects one of the output classes as the winning one for every timestep of an utterance, and the most voted class during the duration of an utterance is selected as the prediction for that particular utterance: this approach is usually referred to as winner-takes-all. 
Since this is a classification task and not a time-series prediction task, we don't use use anymore the NMSE as a figure of merit but the Error Rate. The Error Rate is simply defined as the ratio between the utterances predicted correctly and the total amount of utterances. 

\subsection{Speaker Recognition}
The Japanese Vowels dataset \cite{kudo1999multidimensional} consists of 640 utterances of the Japanese vowel ‘ae’, pronounced by 9 individuals. The goal is to correctly identify the speaker for each utterance. To keep consistency with previous works, the database is split in 270 train sequences and 370 test sequences. The audio samples are pre-processed using the Mel-frequency cepstral coefficients (MFCCs) to obtain their frequency representation, used as input signal to the reservoir. Each sample consists thus of 12 MFCC coefficients for every timestep. The length of the utterances ranges from 7 to 29 timesteps, with an average length of approximately 15 timesteps each. For statistical purpose, the training procedure is repeated 10 times: each time we select at random 270 audio samples for the training and use the rest for testing.  Also here, as with the Spoken Digit Recognition task, the Error Rate and the winner-takes-all approach are used to evaluate the accuracy of the system.

\section{Results}
\label{sec_results}

In this section we show the experimental results obtained with our optoelectronic reservoir on the tasks listed in section \ref{sec_tasks}. We use $N=50$ reservoir nodes for NARMA10 and Mackey-Glass, and $N=100$ nodes for the audio tasks. For the NARMA10 and Mackey-Glass tasks, we use a time-series length of $K=21000$. We divide these data points in four sets: the first 500 points are removed for washout, the next 10000 points are used for training, then 500 points are again removed for washout, and the last 10000 points are used for testing.
For washout we mean the removing of initial or transient states in a system, to ensure that the model focuses on the system's steady-state behavior rather than transient effects that might not represent the system's long-term dynamics.
For the Digits and the Speakers tasks, the input length is given by the dataset itself (cf. section \ref{sec_tasks}). The frequency-encoded audio samples are sent to the reservoir sequentially: the state of the reservoir are not reset between every digit, meaning that the delayed input can combine data points coming from the present and the previous digit.

To test the effectiveness of the optimisation using the delayed input, we use the same approach for all the tasks. 
The input scaling $\beta_1$ and the bias $J_0$ (cf. equation (\ref{eq_dly_inp})) are selected for each task so that the range of the input $J(n)$ falls in the range [0.2, 1.4] when $\beta_2$ is set to zero. These values, though not optimized, are considered reasonable inputs for the reservoir. The feedback attenuation is set with the OA of Figure \ref{fig_setup} to 2 dB (which roughly corresponds to a $\alpha=0.15$ in equation (\ref{eq_T-M})), a reasonable operating value for our optoelectronic reservoir, and the ridge regression parameter is set to $\lambda=10^{-5}$ for all the tasks. We emphasize that we do not optimise the  value of $\alpha$, $\beta_1$ and $\lambda$ as hyperparameters for every task, as it is usually done in RC. Another important parameter that is not optimised is the feedback delay time.  In fact, in this work we want to highlight the reliability of using the delayed input for optimisation without having to rely on the traditional tuning of hyperparameters. Moreover, we test the NARMA10 and Mackey-Glass tasks also with a value of feedback parameter of 15 dB (which roughly corresponds to $\alpha = 10^{-4}$), very far from reasonable values, to assess the effectiveness of this approach when the hyperparameters are off from acceptable values.  Table \ref{tab_params} contains in a more compact form all the values above mentioned, for all the tasks.

\begin{table}[h!]
\centering
\begin{tabular}{| c || c c c c c||} 
 \hline
  & $\beta_1$ & $J_0$ & Opt. att. & $\lambda$ & $N$   \\ 
 \hline\hline
 NARMA10 & 1.8 & 0.4 & 2 dB \& 15 dB& $10^{-5}$ & 50 \\ 
 Mackey-Glass & 1  & 0 & 2 dB \& 15 dB& $10^{-5}$ & 50  \\
 Spoken Digit & 10 & 0 & 2 dB & $10^{-5}$ & 100 \\
 Speaker & 0.3 & 0.8 & 2 dB & $10^{-5}$ & 100 \\
 \hline
\end{tabular}
\caption{Parameters of the input and the reservoir, for all the considered tasks. The optical attenuation (opt. att.) is set by the OA in Figure \ref{fig_setup}, and is related to the hyperparameter $\alpha$ of equation (\ref{eq_T-M}).}
\label{tab_params}
\end{table}

The only two parameters that we need to tune for each task are then the scaling of the delayed part of the input $\beta_2$ and the delay $d$ (cf. equation (\ref{eq_dly_inp})). We include in the search also the case where $\beta_2$ is scanned while $d = 0$, which gives a rough idea of how the reservoir behaves when scanning the input strength with no delayed input (cf. section \ref{sec_dly_vs_std} for a more detailed comparison of the optimisation with and without delayed input).  

\subsection{Effectiveness of optimisation with delayed input}
\label{sec_results_Effectiveness}

\begin{figure}[h!]
\centering\includegraphics[width=14cm]{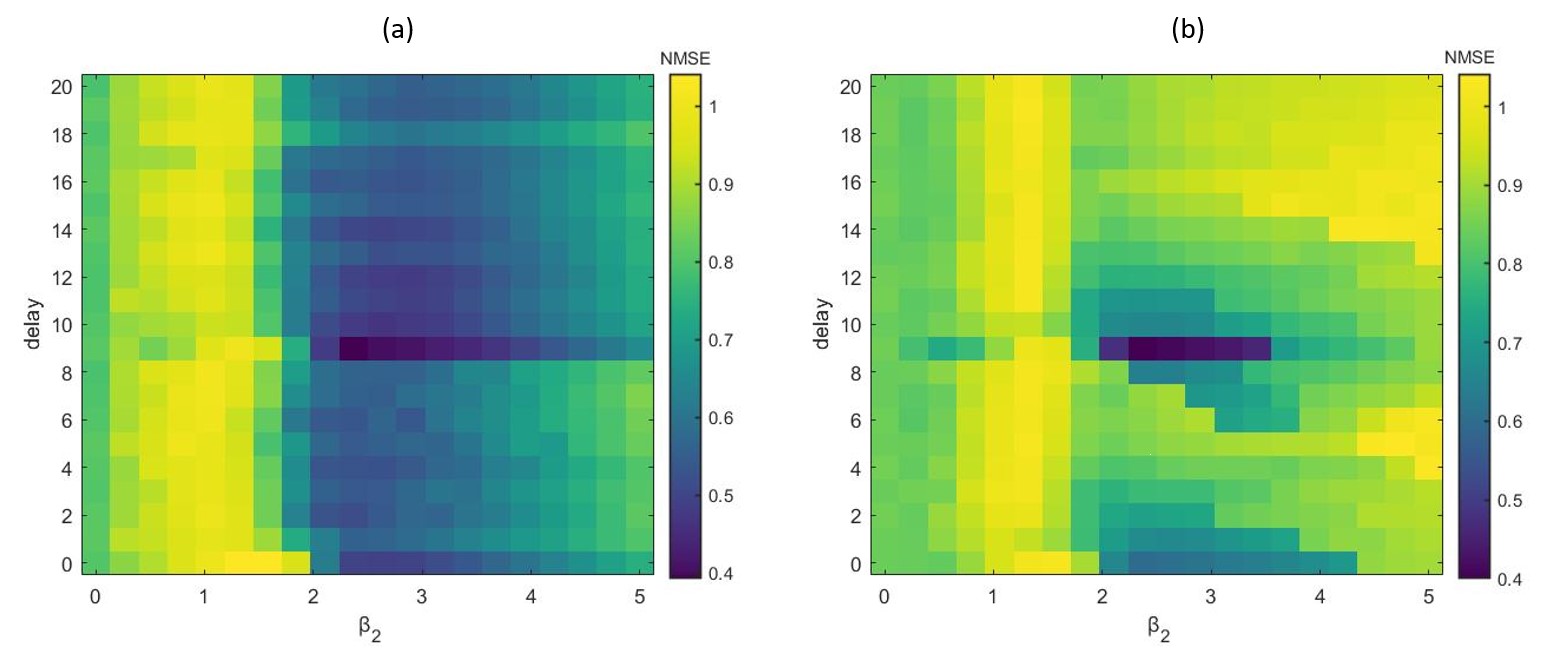}
\caption{Experimental results on the NARMA10 task as a function of $\beta_{2}$ and the delay $d$ of equation (\ref{eq_dly_inp}), using different optical feedback attenuation: 2 dB (a) and 15 dB (b).}
\label{fig_narma}
\end{figure}

Figure \ref{fig_narma} show results on the NARMA10 task. We scan the parameters $\beta_2$ and $d$ looking for the best configuration. 
Our best results is a NMSE of 0.4: it corresponds to a delay of 9 timesteps, most likely due to the intrinsic timescales of the NARMA10 task (cf. equation (\ref{eq_narma})). As a comparison, in \cite{hermans2016embodiment} authors reach a best NMSE of 0.32 with a similar optoelectronic system, in all likelihood because of the higher number of virtual nodes used ($N=80$ while we use $N=50$). Moreover, the results in \cite{hermans2016embodiment} are obtained through a more advanced training method based on physical backpropagation, while we use a simple scanning of two parameters.
In \cite{antonik2016online} authors use an FPGA-based optoelectronic reservoir similar to ours and reach a NMSE of 0.2, but using a more complex training scheme based on an analog output layer and an online learning algorithm. 
In \cite{Jaurige2021reservoir} authors reach a NMSE of 0.3 using the delayed input approach in numerical simulations. 

Additionally, we wanted to test how the optimisation with delayed input works in the case where the reservoir hyperparameters are rather poor.
Figure \ref{fig_narma}(a) and Figure \ref{fig_narma}(b) show results corresponding to reservoirs where a different feedback attenuation $\alpha$ is applied: the attenuation is 2 dB in Figure \ref{fig_narma}(a) and 15 dB in Figure \ref{fig_narma}(b). As explained previously, with 2 dB our setup can work reasonably on various tasks; instead, 15 dB is a high attenuation value that brings the system in a operating region of poor performance. Nonetheless, it is clear from Figure \ref{fig_narma}(b) that using a delayed input, for this task, allows to reach the best performance even while operating in this sub-optimal hyperparameter region. We also find quite poor performance without delay ($\beta_2 = 0$) in both cases, most likely because the hyperparameters of the reservoir (such as $\alpha$, $\beta_1$, $\lambda$) are not optimised.

\begin{figure}[h!]
\centering\includegraphics[width=14cm]{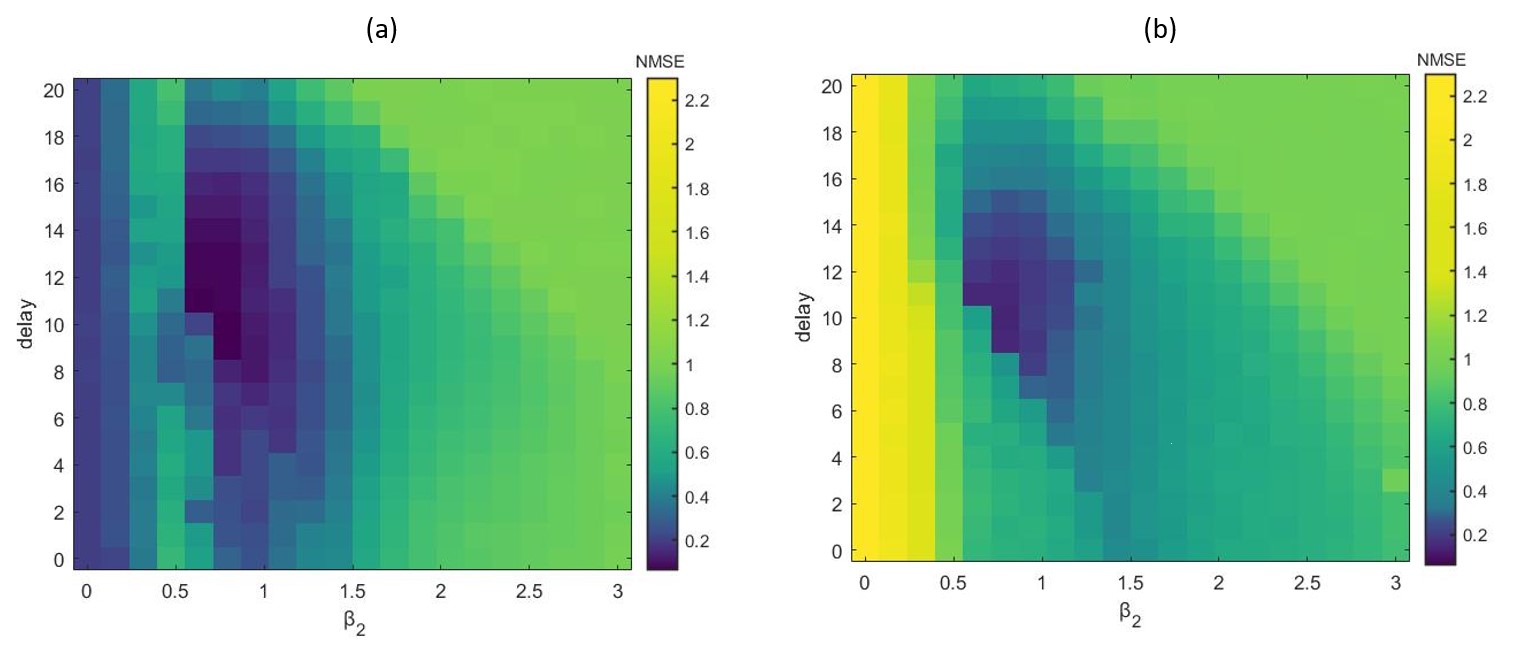}
\caption{Experimental results on the prediction of Mackey-Glass system 10 steps in the future, as a function of $\beta_{2}$ and the delay $d$ of equation (\ref{eq_dly_inp}), using different optical feedback attenuation: 2 dB (a) and 15 dB (b).}
\label{fig_mg}
\end{figure}

Similar results, shown in Figure \ref{fig_mg}(a), are obtained on the prediction of the Mackey-Glass system. Without delay ($\beta_2=0)$ the performance are already acceptable (NMSE = 0.20), but they get further improved with the delay, reaching an NMSE = 0.063 with $d=11$.
Similar results were obtained on the same task in numerical simulations in \cite{Jaurige2021reservoir}, using the delayed input approach. Our results are also comparable to previous works \cite{ortin2015unified, antonik2018using} which use similar time-delay reservoirs with thousands of nodes.
In the case of an attenuation of 15 dB (Figure \ref{fig_mg}(b)), without delay the system performs very poorly, with the NMSE exceeding 2; instead, by tuning the delayed input we reach an NMSE = 0.141. This proves that, even when very inadequate hyperparameters are selected, the use of a delayed input can still bring the reservoir in a valuable operating region.
Note that the optimal delay $d$ is different from the delay $\tau$ in  equation (\ref{eq_MG}) (while one would naively suppose they should be similar).


\begin{figure}[h!]
\centering\includegraphics[width=9.5cm]{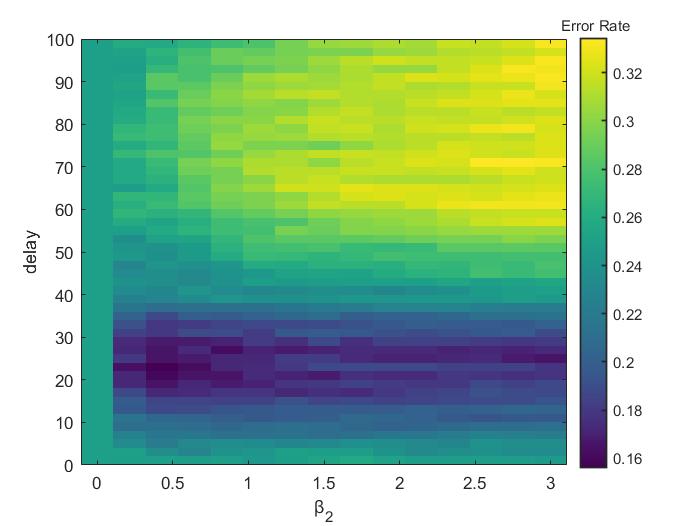}
\caption{Experimental results on Spoken Digit Recognition with noise, as a function of $\beta_{2}$ and the delay $d$ of equation (\ref{eq_dly_inp}).}
\label{fig_digits}
\end{figure}

The results on "noisy" Spoken Digit Recognition are reported in Figure \ref{fig_digits}. For this task the error rate is considerably improved using the delayed input, decreasing from 0.250 with no delay ($\beta_2=0$) to 0.156 with a delay $d=23$. In \cite{picco2023deep}, on the same task and with the same optoelectronic setup, we reported similar performance but using a deep architecture with 2 interconnected reservoir of size $N=100$ each, or using a single reservoir of size $N=600$, while in this work we use only one reservoir of size $N=100$.   

\begin{figure}[h!]
\centering\includegraphics[width=10cm]{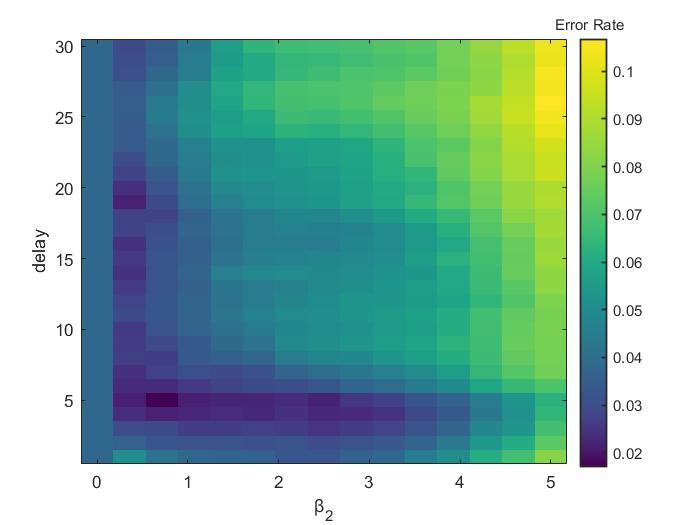}
\caption{Experimental results on Speaker Recognition, as a function of $\beta_{2}$ and the delay $d$ of equation (\ref{eq_dly_inp}). }
\label{fig_speakers}
\end{figure}

Figure \ref{fig_speakers} shows results on the Speaker Recognition. Also in this case the use of a delayed input increase the performance. The error rate decreases from 0.0384 with no delay ($\beta_2=0$) to 0.0170 error with a delay $d=5$. As a comparison, in \cite{paudel2020classification} authors reach a best error of roughly 0.020 with an optical reservoir of $N=200$ nodes, whereas in \cite{dale2023reservoir} authors report an error of roughly 0.035 with a reservoir size comparable to ours.

For these two latter tasks involving audio processing, the explanation for the optimum values of the delay is not straightforward. For both tasks, the optimum delay value is shorter than the average duration of an utterance: the Digits task has an average sample length of 68 timesteps and an optimum delay $d=23$, while the Speakers task has an average sample length of 15 timesteps and an optimum delay $d=5$. A delay shorter than the utterance duration means that most of the times the delayed input term and the non-delayed input term refer to the same utterance. Thus, these results suggest that the performance of the reservoir can be increased when different temporal components of the same utterance are combined. 
While $d$ and the sample duration seem somewhat correlated, their relationship does not look immediately quantifiable and it is not possible so far to estimate the optimum delay a priori. This may be due to the different audio preprocessing algorithms used for the two datasets. 

\subsection{Optimisation with delayed input vs standard hyperparameters optimisation}
\label{sec_dly_vs_std}
\begin{figure}[h!]
\centering\includegraphics[width=14cm]{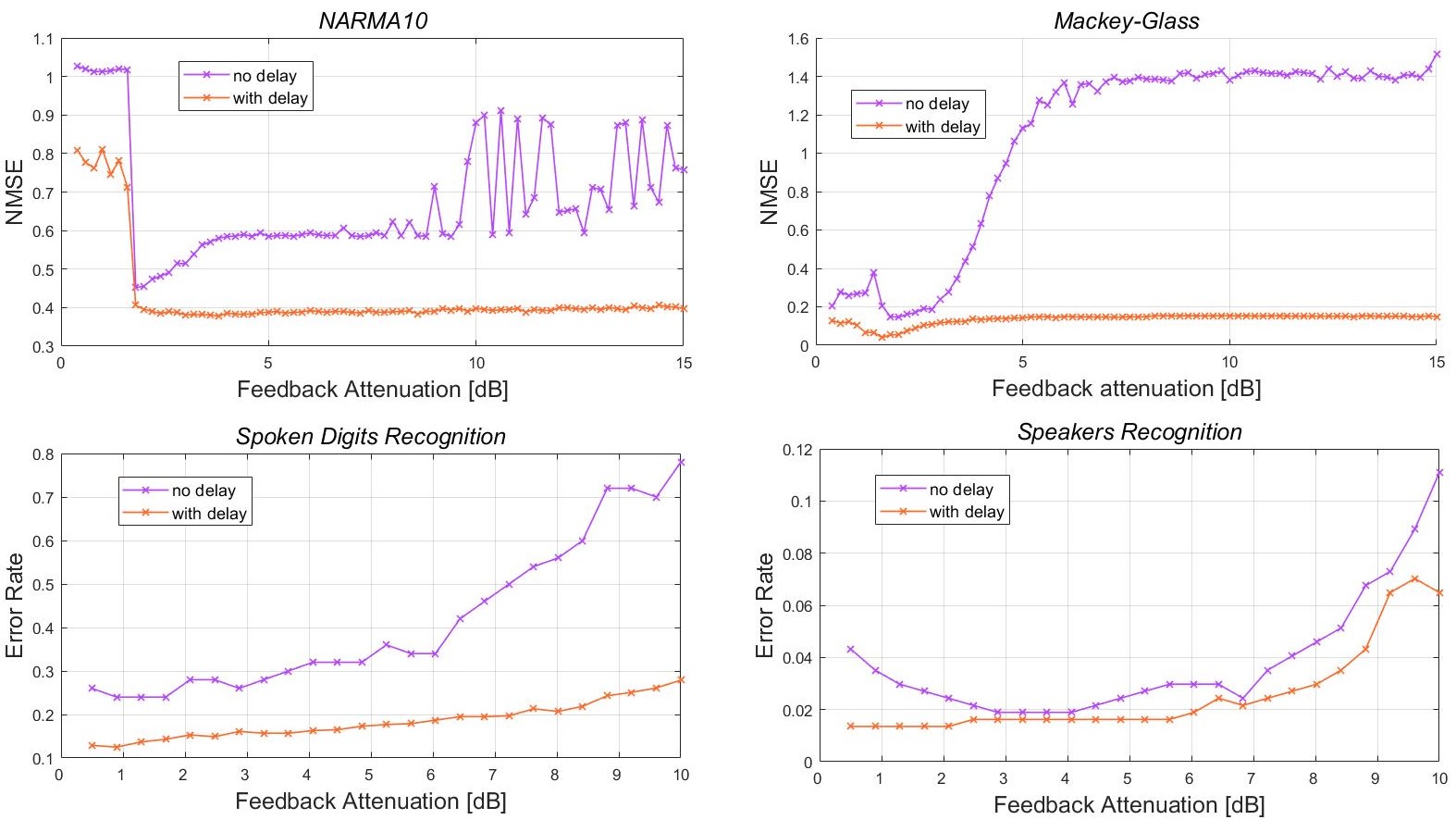}
\caption{Experimental results on all the tasks, to compare the standard approach (without delay) and the approach proposed in this work (with delay). The purple curves are obtained using no delay, optimising the hyperparameters $\beta_1$ and $\lambda$ and then sweeping the optical feedback attenuation. The orange curves are obtained using the delayed input, optimising the parameters $\beta_2$ and $d$ of equation (\ref{eq_dly_inp}), and then sweeping the optical feedback attenuation. }
\label{fig_vsalpha}
\end{figure}

The previous section reports how, for all the tasks, the use of a delayed input outperforms the case where no delay is used, i.e. $\beta_2=0$. However, all the results presented so far refer to a reservoir whose hyperparameters are not optimised, but rather set to a reasonable value.
What happens instead if the hyperparameters are optimized for each task ? Would the superiority of the optimisation with delayed input still hold true?
We try to answer this question with Figure \ref{fig_vsalpha}. The purple curves represent the "standard" approach in RC with no delay, i.e. tuning the hyperparameters (in this case $\alpha$, $\beta_1$ and $\lambda$) for every task. The orange curves represent the optimisation using the delayed input, by setting the parameters to the reasonable values of Table \ref{tab_params} and tuning only the delay parameters $\beta_2$ and $d$. In both cases, we explore the behaviour of the system when sweeping the optical attenuation (expressed in dB), which is related to the feedback attenuation $\alpha$.

For all the tasks, it is clear that the optimisation with a delayed input prevails: its accuracy is higher for every value of attenuation considered. Especially for an optical attenuation larger than 10 dB, where performance without delay are very bad, the use of a delay ensures accuracy remains high. 

\section{Conclusions}
In this work we investigate a novel approach for the optimisation of physical reservoir computers using a delayed input, introduced numerically in \cite{Jaurige2021reservoir}. We verify its effectiveness using an optoelectronic setup on different benchmark tasks, involving time-series prediction and audio signal classification. 

We emphasize the advantages of this approach. 
It is a very simple technique that does not require complex search algorithms. In fact, the reservoir's performance can be optimised only by tuning two parameters: the strength of the delayed component ($\beta_2$) of the input, and the delay itself ($d$). Therefore, it is not necessary to carefully select the reservoir's hyperparameters for every task: it is sufficient to set them to some reasonable value and fine tune the delayed input. In addition, the performance is not very sensitive on the value of $\beta_2$: values in the range 1-3 are good, and need a bit of refinement to get the best values. Since one only needs to act on the input signal, this approach can be useful for physical reservoirs whose hyperparameters cannot be tuned due to real-world constraints.
Furthermore, even in the case where the hyperparameters are considerably off from acceptable values, the use of a delayed input can still bring the reservoir to satisfactory performance. Moreover, this optimisation procedure outperforms the traditional hyperparameters tuning in terms of performance. 

Some open questions which we leave open for future research concern the relationship between the timescale of a temporal task and the optimum amount of delay to employ; the use of multiple delays (in which case improved optimisation schemes, such as Bayesian optimisation, rather than the simple grid scan used here, would be useful); and on the experimental side the implementation of the delayed input with a physical delay line.


\begin{backmatter}
\bmsection{Funding}
E.P and S.M. acknowledge financial support from the H2020 Marie Skłodowska-Curie Actions (Project POSTDIGITAL Grant number 860360), and from the Fonds de la Recherche Scientifique - FNRS, Belgium under funding EOS n° 40007536 and CDR n° 40021243. L.J. and K.L. acknowledge funding from the Deutsche 369 Forschungsgemeinschaft (DFG), Grant No. LU 1729/3-1 370 (Projektnummer 445183921).


\end{backmatter}

\appendix
\bibliography{sample}






\end{document}